\documentclass[aps,prb,twocolumn]{revtex4-1}
\usepackage{graphicx}
\usepackage{color}
\usepackage[latin1]{inputenc}
\usepackage{amsbsy,amsmath,amssymb,amsthm,amsfonts}
\usepackage{times,mathptmx}
\usepackage{bm}
\usepackage{units}

\newcommand{\kk}{{\bf k}}
\newcommand{\qq}{{\bf q}}

\newcommand{\braket}[1]{\langle #1 \rangle}

\begin{document}
\title{Impact of doping on the carrier dynamics in graphene}

\author{Faris Kadi$^{1}$}
\email{ermin.malic@chalmers.se}
\author{Torben Winzer$^{1}$}
\author{Andreas Knorr$^{1}$}
\author{Ermin Malic$^{2}$}
\affiliation{$^{1}$Institut f\"ur Theoretische Physik, Technische Universit\"at Berlin,
  Hardenbergstr. 36, 10623 Berlin, Germany}
\affiliation{$^{2}$Chalmers University of Technology, Department of Applied Physics, SE-412 96 Gothenburg, Sweden}

\begin{abstract}

We present a microscopic study on the impact of doping on the carrier dynamics in 
graphene, in particular focusing on its influence on the technologically relevant carrier multiplication in realistic, doped graphene samples.
Treating the time- and momentum-resolved carrier-light, carrier-carrier, and carrier-phonon interactions
on the same microscopic footing, the appearance of Auger-induced carrier multiplication up to a Fermi level of \unit[300]{meV} is revealed. Furthermore, we show that doping favors the so-called hot carrier multiplication occurring within one band. Our results are directly compared to recent time-resolved ARPES measurements and exhibit an excellent agreement on the temporal evolution of the hot carrier multiplication for n- and p-doped graphene.
The gained insights shed light on the ultrafast carrier dynamics in realistic, doped graphene samples.
\end{abstract}
\maketitle
\section{Introduction}
A number of theoretical and experimental studies has been performed aiming at a thorough understanding of the carrier relaxation dynamics in 
optically excited graphene.\cite{erminbuch,dawlaty08,plochocka09,wang10,obraztsov11,winnerl11,sun12,heinz13-3,johannsen13,gierz13,kadi14, winnerl14, winzer15, winnerl15}
Most of these studies focus on the ultrafast Coulomb- and phonon-induced carrier dynamics without considering the influence of doping in the investigated graphene samples.
A non-zero Fermi level can have a crucial impact on the relaxation dynamics via a significant increase of the scattering phase space and via the enhancement of Pauli blocking. A first experimental time-resolved ARPES study has been performed addressing the doping dependence of carrier multiplication of graphene.\cite{hofmann15} The underlying elementary processes determining the observed different behavior for p- and n-doped samples have not been microscopically investigated, yet. 

In this work,  we apply a microscopic approach to access the time-, momentum-, and angle-resolved dynamics of electrons and phonons in optically excited graphene under the influence of a variable n- and p-doping.
The focus lies in particular on the impact of a finite Fermi level on the appearance of the technologically relevant carrier
multiplication.\cite{winzer10, song11, winzer12-2, pirro12, tielrooij13, brida13, song13, basko13, ploetzing14, wendler14, gierz15, hofmann15}

This interesting ultrafast phenomenon is related to the linear electronic band structure of graphene opening up the possibility of efficient Coulomb-induced Auger processes. A significant multiple carrier generation has been theoretically predicted\cite{winzer10, winzer12-2, wendler14} and experimentally confirmed in 
graphene.\cite{brida13, tielrooij13,  ploetzing14, hofmann15, gierz15}
So far, the theoretical studies have been constrained to the case of undoped graphene. Introducing a non-zero Fermi level in graphene, electrons above the Dirac point or above the Fermi level can be considered as charge carriers (holes in analogy). In the first case, the carrier multiplication can take place via Auger scattering bridging the valence and the conduction band. In the following, we label this process as \textit{carrier multiplication (CM)}. On the other side, counting carriers with respect to the Fermi level, the multiplication occurs via Coulomb-induced intraband scattering bridging the states below and above the Fermi level, cf. Fig. \ref{fig1}. According to  literature,\cite{tielrooij13} we label this process as \textit{hot carrier multiplication (hCM)}. Here, the actual number of charge carriers remains unchanged in each band. Nevertheless, since these hot carriers are crucial for many technological applications, the appearance of hCM is also of technological relevance.

\section{Theoretical approach} 

The starting point for the calculation is the many-particle Hamilton operator $H = H_0 + H_{c,f} + H_{c,p} + H_{c,c}$, where $H_0$ 
denotes the interaction-free carrier and phonon part, 
 $H_{c,l}$ the carrier-light coupling, 
 $H_{c,p}$ the carrier-phonon interaction, and
$H_{c,c}$ the carrier-carrier interaction.\cite{haug04} 
The carrier dynamics is described by graphene Bloch equations\cite{erminbuch} corresponding to a coupled set of differential equations for the occupation probability $\rho_{\bf k}^\lambda(t)$ in the state ${\bf k}$ and the
band $\lambda=(v,c)$, the microscopic polarization $p_{\bf k}(t)$ that is a measure for the optical transition probability between both bands, and the phonon occupation $n^j_{{\bf q}}(t)$ with the momentum ${\bf q}$ for different 
optical and acoustic phonon modes $j$
\cite{malic11-1}:

\begin{align}
\label{eq:pol} \dot{p}_{{\bf k}}(t) =&  \Big( i\omega^{vc}_{\bf k} + i[\Omega^{vv}_{\bf k}(t)-\Omega^{cc}_{\bf k}(t)] -\gamma_{\bf k}\Big) p_{\bf k}(t) \\
\nonumber &+i \Omega^{vc}_{\bf k} (t)\Big(\rho^v_{\bf k}(t)-\rho^c_{\bf k}(t)\Big) p_{\bf k}(t) + {\cal U}_{\bf k}(t) {\rm ,}\\[12pt]
\label{eq:rhoe} \dot{\rho}_{{\bf k}}^{\lambda} (t) =&\pm 2 \Im \Big(\Omega^{vc*}_{\bf k}(t)p_{\bf k}(t)\Big)\\
\nonumber &+ \Gamma^{\text{in}}_{\lambda,{\bf k}}(t)\Big(1-\rho_{{\bf k}}^{\lambda}(t)\Big)-\Gamma^{\text{out}}_{\lambda,{\bf k}}(t)\rho_{{\bf k}}^{\lambda}(t) {\rm ,}\\[12pt]
\label{eq:nq} 
\dot{n}^j_{{\bf q}}(t)=&-\gamma_{\text{ph}}\Big(n^j_{{\bf q}}(t)-n^j_B\Big)\\
\nonumber &+\Gamma^{\text{em}}_{j,{\bf q}}(t)\Big(1+n^j_{{\bf q}}(t)\Big)-\Gamma^{\text{abs}}_{j,{\bf q}}(t)n^j_{{\bf q}}(t).\\\nonumber
\end{align}
Here, $\omega^{vc}_{\bf k}=\frac 1 \hbar (\varepsilon^{v}_{\bf k}-\varepsilon^{c}_{\bf k})$ is the optical transition 
frequency within the linear electronic band structure $\varepsilon^{\lambda}_{\bf k}$ of graphene close to the Dirac point. The carrier-light coupling is determined by 
  $\Omega^{\lambda\lambda'}_{\bf k}(t)=i\frac{e_0}{m_0}{\bf M}_{\bf k}^{\lambda \lambda'}\cdot{\bf A}(t)$ with the optical matrix element\cite{malic11-1} ${\bf M}_{\bf k}^{\lambda \lambda'}$, the vector 
  potential ${\bf A}(t)$ representing the
 excitation pulse, the free electron mass $m_0$, and the charge $e_0$. Here, $\Omega^{vc}_{\bf k}(t)$ is the Rabi frequency and $\Omega^{\lambda\lambda}_{\bf k}(t)$ accounts for intraband transitions.\cite{kadi14}
The many-particle interactions 
are treated within the second-order Born-Markov approximation,\cite{haug04,erminbuch,knorr96,kochbuch11} which yields
a Boltzmann-like scattering equation for the carrier occupation with the time- and momentum-dependent scattering rates $\Gamma_{\kk,\lambda}^{\text{in,out}}(t)$ accounting for Coulomb- and phonon-induced processes.
At the same time, the microscopic polarization is damped by the many-particle-induced diagonal dephasing $\gamma_{\bf k}(t)$ and is driven by the off-diagonal dephasing term ${\cal U}_{\bf k}(t)$.
In analogy, the equation of motion for the phonon occupation $n^j_{{\bf q}}(t)$ is obtained and contains phonon emission and absorption rates $\Gamma^{\text{em,abs}}_{j,{\bf q}}(t)$.
The finite phonon lifetime\cite{kang10} $\gamma_{\text{ph}}$ is considered by a coupling to a phonon bath $n^j_B$ at room temperature.
The explicit form of the time- and momentum-dependent scattering rates is discussed in the supplementary material. More details on the diagonal and off-diagonal dephasing terms can be found in Malic et al.\cite{malic11-1}
\begin{figure}[t!]
  \begin{center}
\includegraphics[width=0.35\linewidth]{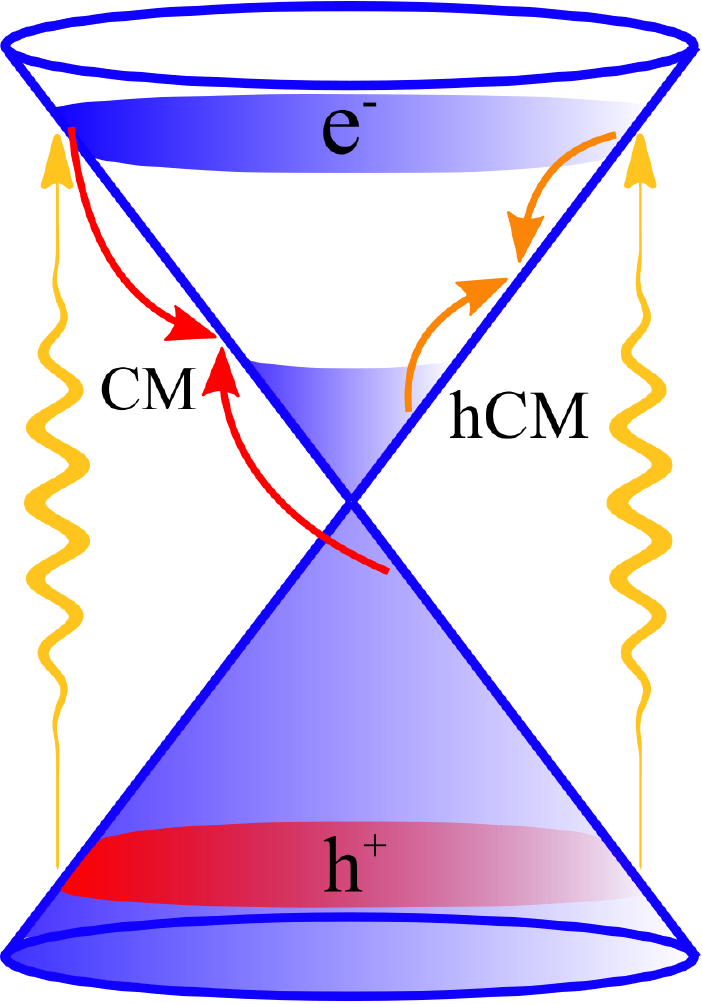}
  \end{center}
  \caption{\textbf{Schematic illustration of Coulomb-induced scattering processes for optically excited n-doped graphene.} In the presence of a finite Fermi level, carrier multiplication (CM) and hot carrier multiplication (hCM) need to be distinguished. While the first is induced by Auger processes bridging the valence and the conduction band (red arrows) and increasing the number of charge carriers in the conduction band, the second corresponds to Coulomb-induced intraband scattering (orange arrows) and increases the number of  carriers above the Fermi level.} \label{fig1} 
\end{figure}

A finite Fermi level $E_F$ breaks the symmetry between the
valence and conduction band around the Dirac point, cp. Fig. \ref{fig1}. As a result, the occupation probability of electrons $\rho_{\bf k}^e=\rho_{\bf k}^c$ and
of holes $\rho_{\bf k}^h=1-\rho_{\bf k}^v$ needs to be treated separately. As initial condition, we assume a Fermi distribution $\rho_{{\bf k}}^{(e,h)}=\left[\exp\left((\varepsilon^{(e,h)}_{\bf k}\pm E_F) / k_B T\right) + 1\right]^{-1},$ where $+$ stands for the hole and $-$ for the electron occupation at the temperature $T$. Another important aspect of doping is the increased screening of the Coulomb interaction. 
The bare Coulomb potential $V_\qq$ appearing in the Coulomb-induced scattering rates $\Gamma_{\kk,\lambda}^{\text{in,out}}(t)$  is screened via the dynamic dielectric
function $\varepsilon(\qq,\omega)$ that is defined by the Lindhard 
equation,\cite{haug04,giuliani05} cf. the supplementary material for a more detailed discussion.
Since this many-particle-induced screening is directly influenced by carrier occupations in the conduction and valence bands, doping plays a crucial role and has a significant influence on the ultrafast
carrier dynamics in graphene.

\begin{figure}[t!]
  \begin{center}
\includegraphics[width=\linewidth]{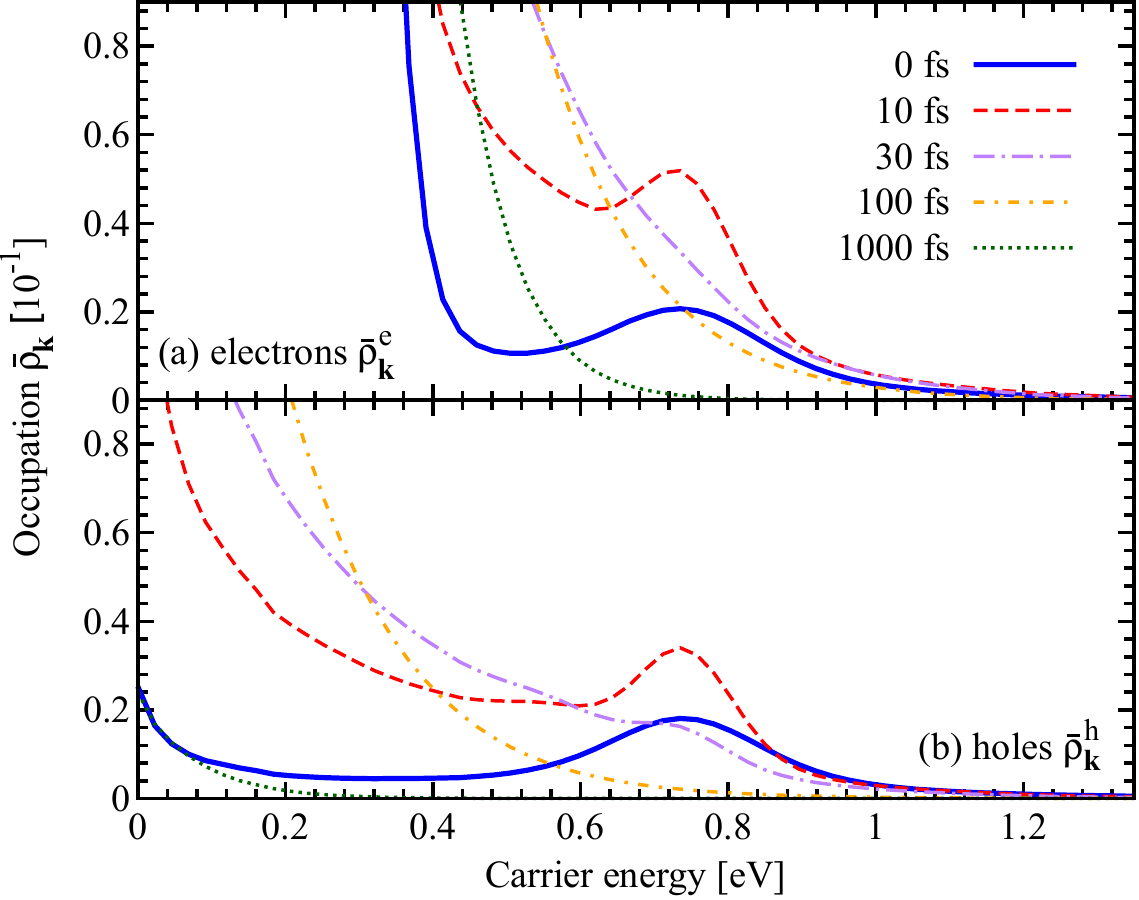}
  \end{center}
  \caption{\textbf{Electron and hole occupations in doped graphene.} Angle-averaged occupation probability $\bar{\rho}_{\bf k}$ for (a) electrons and (b) holes in highly 
  n-doped graphene ($E_F=\unit[300]{meV}$) is shown as a function of the carrier energy for different times after the optical excitation. Note that due to the n-doping the electron relaxation is faster resulting in a  thermalized hot Fermi distribution already after \unit[30]{fs}. The inverse behavior can be found for p-doped graphene.
   } \label{fig2} 
\end{figure}

With the presented microscopic approach, we can track the relaxation dynamics of non-equilibrium charge carriers in time, and energy including the temporal
evolution of the carrier density after the optical excitation. 
First, we focus on the electron and hole dynamics in highly doped graphene and then we discuss the impact of doping on the appearance of carrier multiplication.

\section{Electron and hole dynamics} 
Here, we discuss how the doping-induced symmetry breaking  between electrons in the conduction band and holes in the valence band influences the dynamics of optically excited charge carriers in realistic doped graphene samples. Figure \ref{fig2} illustrates the angle-averaged occupation probability $\bar{\rho}^\lambda_{\bf k}$ for (a) electrons and (b) holes for an initial Fermi level of \unit[300]{meV} as a function of 
the carrier energy for different times after the optical excitation. Note that for symmetry reasons, the physical picture remains the same in p-doped graphene samples but electrons and holes switch their roles, respectively. The system is excited by a \unit[10]{fs} 
pulse with a photon energy of \unit[1.5]{eV} and a pump fluence of $\unit[0.3]{\mu J cm^{-2}}$. The characteristics of the excitation pulse correspond to typical values that can be realized by standard pulsed lasers.\cite{breusing11} 
The pulse is centered at \unit[0]{fs} and gives rise to a well pronounced non-equilibrium distribution for electrons and holes around the carrier energy of \unit[0.75]{eV}, cf. Fig. \ref{fig2}.
For both electrons and holes, the efficient carrier-carrier and carrier-phonon scattering leads to an ultrafast thermalization of the system towards a hot Fermi distribution already 
after some tens of femtoseconds. Then, a slower phonon-induced carrier cooling occurs that drives the electron and hole occupations towards their initial thermal Fermi distributions. 
Due to the increased number of available scattering partners in the conduction band of n-doped graphene, the Coulomb-driven carrier thermalization occurs faster for electrons. Here, a hot thermalized Fermi distribution is already reached after \unit[30]{fs}, while at the same time the holes exhibit still a non-equilibrium distribution, cf. the purple lines in Fig. \ref{fig2}.

\section{Carrier multiplication (CM)}
Now, we study the impact of an initial Fermi level $E_F$ on the Coulomb-induced multiple carrier generation, which is generally defined as the ratio between the number of overall generated electron-hole pairs and the optically excited charge carriers
\begin{align}
(h)CM=\frac{n-n_T}{n_{opt}}, \label{eq:cmdef}
\end{align}
where $n$ is the total carrier density, $n_T$ the initial thermal carrier background, and $n_{opt}$ the optically excited carrier density. All contributions contain both electrons  in the conduction band 
as well as holes  in the valence band.
For doped graphene, the specific definition depends on the physical situation: For optical measurements probing vertical carrier transitions, a definition with respect to 
to the Dirac point is reasonable, i.e. $\rho_{\bf k}^e=\rho_{\bf k}^c$ and $\rho_{\bf k}^h=1-\rho_{\bf k}^v$.
\begin{figure}[t!]
  \begin{center}
\includegraphics[width=\linewidth]{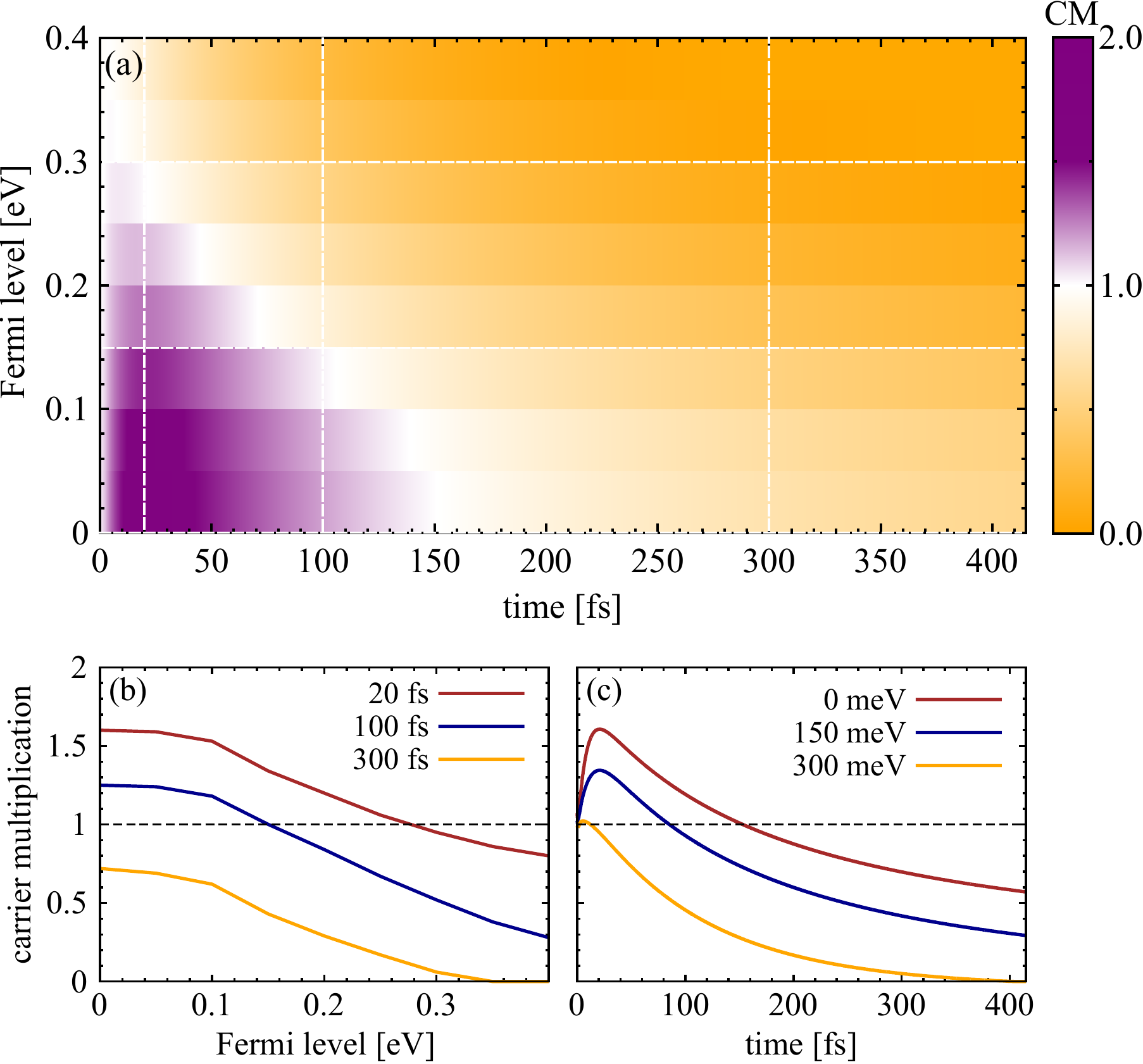}
  \end{center}
  \caption{\textbf{Doping dependence of carrier multiplication (CM).} (a) Temporal evolution of the doping-dependent CM for a fixed absorbed pump fluence of $\varepsilon_{\text{abs}}=\unit[0.3]{\mu J cm^{-2}}$ and an excitation energy of $1.5$ eV. 
  (b) CM as a function of the Fermi level $E_F$ for three fixed time delays and (c) the temporal evolution of CM for three fixed Fermi levels. Note that CM only takes place for $E_F$ smaller than \unit[0.3]{eV} and for times up to \unit[150]{fs}.} \label{fig3} 
\end{figure}
The carrier density then reads
\begin{align}
 n = \frac{\sigma_s \sigma_v}{L^2} \sum_{\lambda=e,h;\,{\bf k}} \rho_{\bf k}^{\lambda}, \label{eq:ndef}
\end{align}
where $L^2$ is the graphene area and $\sigma_s$ ($\sigma_v$) denotes the spin (valley) degeneracy. On the other side, electric transport phenomena are also of great interest, where hot electrons around the Fermi level are relevant giving rise to a hot carrier multiplication, cf. Fig. \ref{fig1}. This situation will be discussed in the next section.

\begin{figure}[t!]
  \begin{center}
\includegraphics[width=\linewidth]{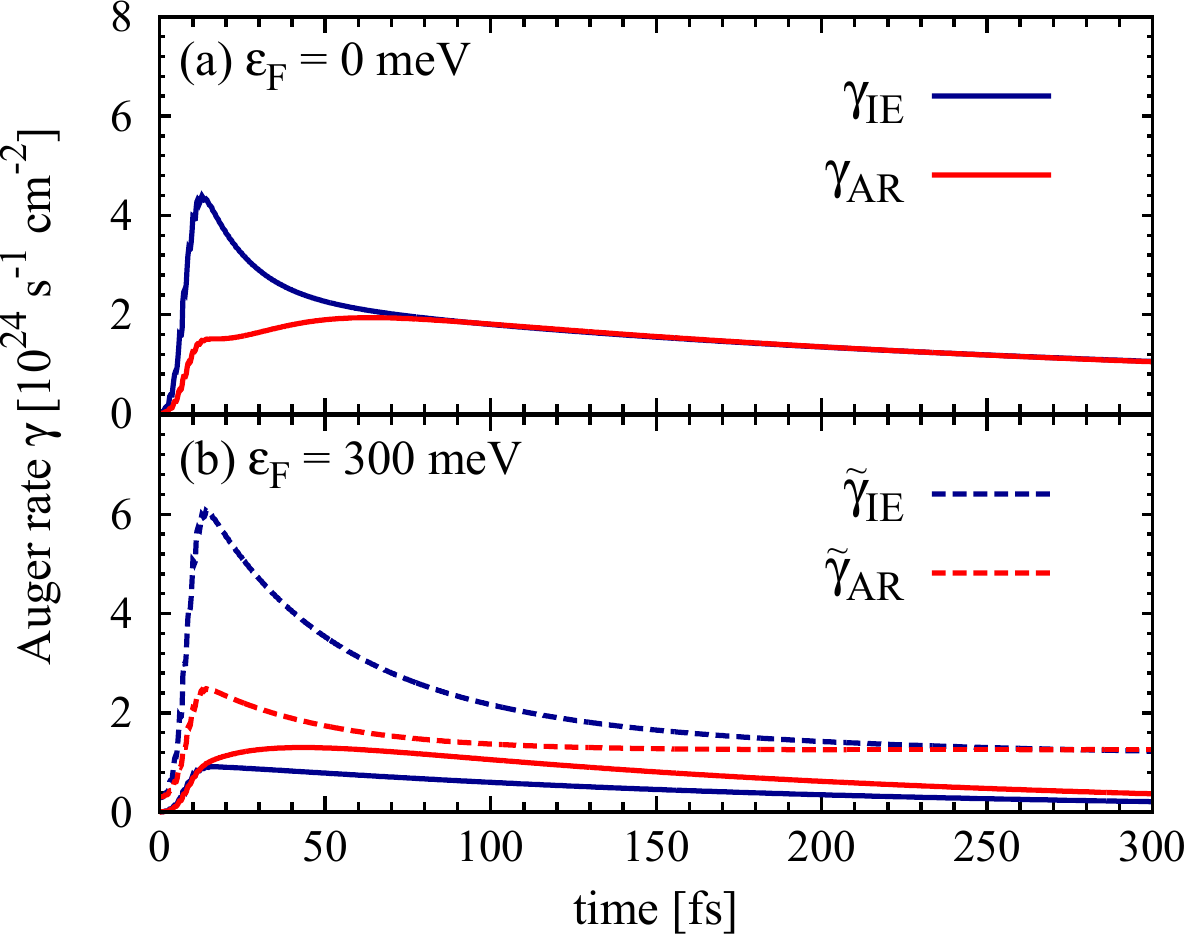}
  \end{center}
  \caption{\textbf{Auger rates in doped graphene.} Temporal evolution of Auger rates including impact excitation (IE) and the inverse process of Auger recombination (AR) for (a) undoped  and (b) highly n-doped graphene ($\varepsilon=\unit[300]{meV}$). 
  In the latter case, also the corresponding \textit{intraband Auger rates} ($\tilde{\gamma}_{IE}$, $\tilde{\gamma}_{AR}$) describing Coulomb-induced scattering bridging the states below and
  above the Fermi level are illustrated (dashed lines), cf. Fig. \ref{fig1}. 
  In undoped graphene, there is a clear asymmetry between both Auger processes in favor of IE in the first \unit[100]{fs}.
  In the doped case,  $\gamma_{AR}$ is significantly larger than $\gamma_{IE}$, however, for \textit{intraband Auger rates}, the situation is
  opposite and  $\tilde{\gamma}_{IE}$ clearly prevails over $\tilde{\gamma}_{AR}$ resulting in a pronounced hot carrier multiplication. } \label{fig4} 
\end{figure}
Treating the full set of graphene Bloch equations,  we have microscopic access to the temporal evolution of the carrier density including the contributions of carrier-light, carrier-carrier, 
and carrier-phonon interactions. 
Figure \ref{fig3}(a) illustrates the temporal evolution of CM as a function of the initial Fermi level $E_F$ at a fixed absorbed pump fluence of $\unit[0.3]{\mu J cm^{-2}}$.  The surface plot reveals that doping clearly reduces the CM efficiency: The lower $E_F$, the higher is the CM factor reaching values of up to approximately $1.7$ 
for undoped graphene at the considered pump fluence, cf. Fig. \ref{fig3}(b). 
CM can be observed for $E_F$  of up to \unit[300]{meV}. Its maximal lifetime is approximately \unit[150]{fs} for undoped graphene and becomes significantly 
shorter for increasing doping, as illustrated in  Fig. \ref{fig3}(c). The observed CM in the low-doping case can be explained by the strongly efficient impact excitation (IE) prevailing over the 
inverse process of Auger recombination (AR), which is a result of the large gradient in carrier occupation around the Dirac point, cf. Fig. \ref{fig2}(a). For undoped graphene, the probability for IE can be written as $\rho^v(1-\rho^c) \approx 1$,
whereas the probability for AR is given by $\rho^c(1-\rho^v) \approx 0$ (since $\rho^v\approx1$ and  $\rho^c\approx0$). In this case, IE is significantly favored by Pauli blocking during the initial dynamics. This is reflected by the corresponding 
rates $\gamma_{IE}$ and $\gamma_{AR}$ that are shown in  Fig. \ref{fig4}(a). We observe that the IE rate is clearly higher
for a time range of approximately \unit[100]{fs} determining the strength  of the appearing CM. During the carrier relaxation both rates converge to the same value and end up in an equilibrium, where no more carriers
are generated. The lifetime of the CM is determined by the duration of the imbalance between IE and AR rates in combination with the interplay with competing channels of carrier-phonon scattering, which transfer energy
from the electronic system to the lattice. With an increasing doping, Auger scattering becomes more and more Pauli blocked resulting in overall lower rates, cf. Fig. \ref{fig4}(b). For Fermi levels 
higher than \unit[300]{meV}, AR becomes the dominant relaxation channels and CM does not appear anymore. 

\begin{figure}[t!]
  \begin{center}
\includegraphics[width=\linewidth]{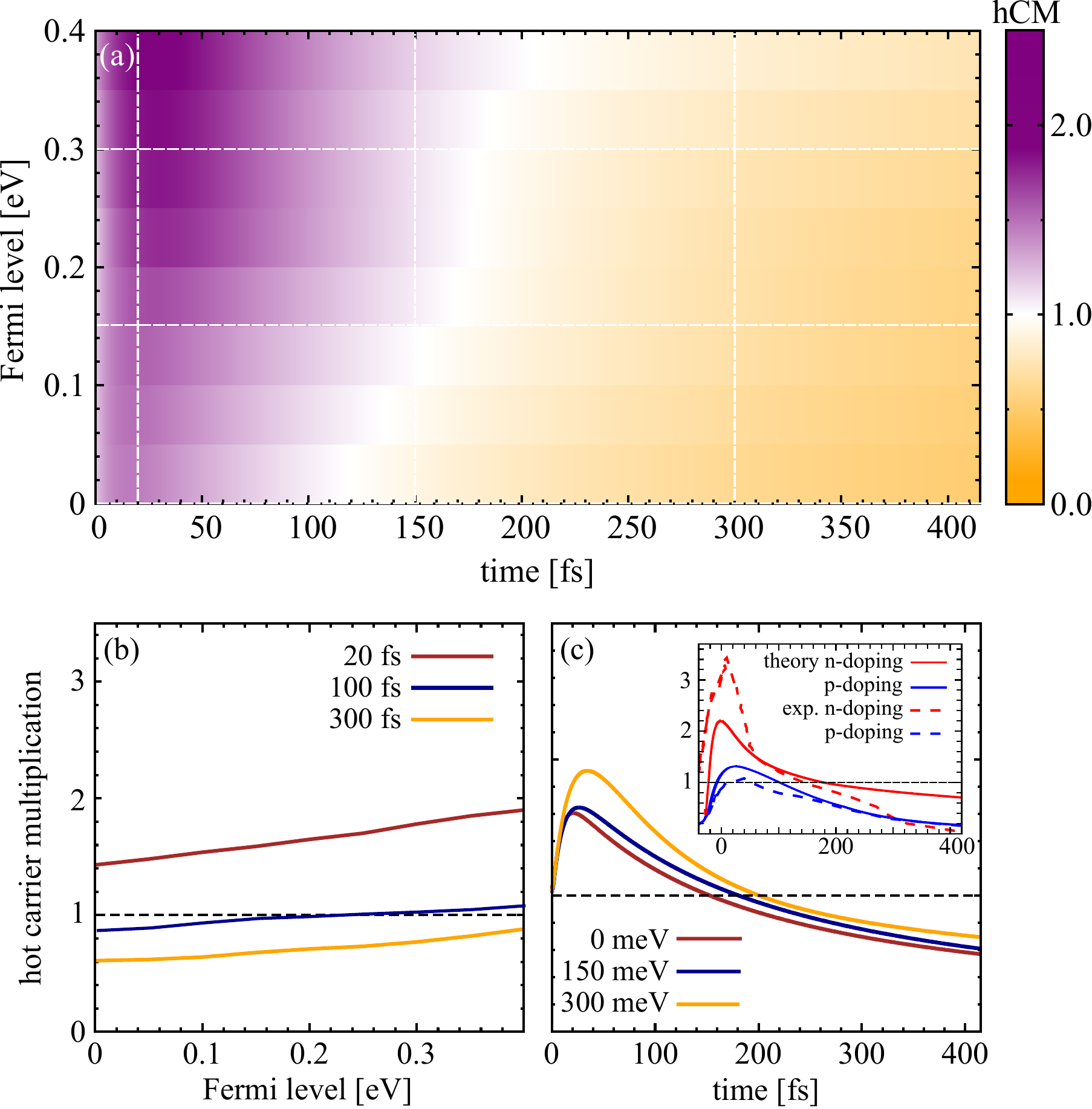}
  \end{center}
  \caption{\textbf{Doping dependence of hot carrier multiplication (hCM).} The same plot as in Fig. \ref{fig4} illustrating now the temporal evolution of the doping-dependent hCM for a fixed absorbed pump fluence of $\varepsilon_{\text{abs}}=\unit[0.3]{\mu J cm^{-2}}$. The inset in (c) shows a direct comparison between theoretically predicted (solid lines)
  and experimentally measured time-resolved ARPES data (dashed lines) for: (i) n-doped graphene ($E_F =$ \unit[380]{meV}, $\varepsilon_{\text{abs}}=\unit[0.5]{\mu J cm^{-2}}$) 
  and (ii) p-doped graphene ($E_F =$ \unit[240]{meV}, $\varepsilon_{\text{abs}}=\unit[1.5]{\mu J cm^{-2}}$). The experimental data is taken from Johannsen et al.\cite{hofmann15}. 
} \label{fig5} 
\end{figure}

\section{Hot carrier multiplication (hCM)} 
Now, we focus on the situation, where charge carriers are defined with 
respect to the Fermi level, i.e. for n-doped graphene the upper Dirac cone is split into $\rho_{{\bf k}}^{e,c}=\rho_{{\bf k}}^c$ for $k>k_F$ and $\rho_{{\bf k}}^{h,c}=1-\rho_{{\bf k}}^c$ for $k<k_F$ with the
Fermi momentum $k_F$. The bottom cone remains unaffected with $\rho_{{\bf k}}^{h,v}\equiv \rho_{{\bf k}}^h=1-\rho_{\bf k}^v$. The according \textit{hot} carrier density $\tilde{n}$ is given by
\begin{align}
 \tilde{n} &= \frac{\sigma_s \sigma_v}{L^2} \left( \sum_{\bf k} \rho_{\bf k}^h+\sum_{\bf k}^{k<k_F} \rho_{\bf k}^{h,c}+\sum_{\bf k}^{k>k_F} \rho_{\bf k}^{e,c} \right),\label{nhCM}
\end{align}
and the associated hot carrier multiplication is highly relevant for transport phenomena,\cite{tielrooij13,hofmann15} cf. Fig. \ref{fig1}.
Note that we obtain symmetric results for n- and p-doped graphene, since the contribution of both electrons and holes to the carrier density is considered. For undoped graphene, both definitions of 
carrier density [Eqs. (\ref{eq:ndef}) and (\ref{nhCM})] and carrier multiplication are equivalent for symmetry reasons.

The surface plot in Fig. \ref{fig5}(a) illustrates hCM as a function of the Fermi level $E_F$ and time at a fixed absorbed pump fluence of $\varepsilon_{\text{abs}}=\unit[0.3]{\mu J cm^{-2}}$.
In contrast to the behavior of CM, we observe a clear increase of hCM with doping.
We reach hCM factors of up to approximately $2$ (at the considered pump fluence) with a lifetime of about \unit[200]{fs} for highly doped 
graphene with $E_F =$ \unit[300]{meV}, cf. Figs. \ref{fig5}(b) and (c). There is nearly a linear dependence between hCM and doping: The smaller $E_F$, the less pronounced is hCM, and the shorter is its lifetime. The probability for intraband IE processes is given by $\rho^{h,c}(1-\rho^{e,c})$, which is initially large compared to 
the probability for intraband AR processes $\rho^{e,c}(1-\rho^{h,c})$, cf. Fig. \ref{fig1}.
With the increasing Fermi level, the \textit{intraband Auger processes} are shifted into a region of higher density of states making them more efficient, as reflected by the much higher scattering rates displayed 
in Fig. \ref{fig4}(b). The initial strong imbalance between  $\tilde{\gamma}_{AR}$ and $\tilde{\gamma}_{IE}$ gives rise to a pronounced hCM. 

Besides the discussed doping dependence, 
CM or hCM are strongly sensitive to the excitation regime. A detailed discussion is provided in the supplementary material, where a semi-analytical approach is presented focusing on the purely Coulomb-induced CM and hCM.

\section{Direct comparison to experimental data} 

After having presented the theoretical results on the doping dependence of the carrier multiplication, we perform a direct comparison with recently performed time-resolved ARPES measurements on n- and p-doped graphene samples.\cite{hofmann15} We explicitly take into account the experimental conditions, such as the Fermi level and the excitation strength. Considering that our microscopic theory does not contain any fitting parameters, we obtain an excellent agreement between theory and experiment, cf. the inset in Fig. \ref{fig5}(c). 
There is a clearly higher hCM for n-doped graphene reaching values of up to $2.2$ in the theory and more than $3$ in the experiment. In contrast, for p-doped graphene only a small hCM of $1.4$ or $1.2$ is obtained in theory and experiment, respectively. This pronounced difference is not due to the type of doping (n, p), as one might assume.\cite{hofmann15} It can be clearly explained by the differences in the applied fluence $\varepsilon_{\text{abs}}$ and the actual Fermi level $E_F$. Note that at the exactly same conditions with respect to $E_F$ and $\varepsilon_{\text{abs}}$, we obtain the same hot carrier multiplication for both n- and p-doped samples. 
However, the experiment has been performed for: (i) n-doped graphene with the Fermi level $E_F =$ \unit[380]{meV} and an absorbed pump fluence of $\varepsilon_{\text{abs}}=\unit[0.5]{\mu J cm^{-2}}$  and (ii) p-doped graphene with $E_F =$ \unit[240]{meV} and $\varepsilon_{\text{abs}}=\unit[1.5]{\mu J cm^{-2}}$. As shown in Fig. \ref{fig5}(b), hot carrier multiplication increases almost linearly with the Fermi level. Furthermore, it is strongly suppressed in the strong excitation regime, i.e. the larger the pump fluence, the less efficient is the hCM, as illustrated in Fig. S3 in the supplementary material.
As a result, the n-doped graphene sample shows a much more pronounced hCM, since its Fermi level $E_F$ is significantly higher and since the experiment has been performed at a clearly smaller pump fluence compared to the p-doped graphene sample.\\

In summary, we have presented a microscopic study of the carrier dynamics in doped graphene samples, in particular focusing on the impact of a finite Fermi level on the (hot) carrier multiplication. We reveal the appearance of Auger-induced carrier multiplication up to Fermi levels of \unit[300]{meV}. In the case of the hot carrier multiplication occurring within one band
doping is even advantageous, since it increases the phase space by providing a large number of available scattering partners. Finally, we have directly compared our results to recent time-resolved 
ARPES measurements finding an excellent agreement and providing a microscopic explanation for the observed different behavior in n- and p-doped graphene samples.
 Our results contribute to a better understanding of the ultrafast carrier dynamics in realistic graphene samples and give valuable insights into the 
technologically relevant carrier multiplication in graphene.

\begin{acknowledgements}
We acknowledge financial support from the Deutsche Forschungsgemeinschaft (DFG)
through SPP 1458 (E.M.), SFB 658 (F.K.) and Sfb 951 (A.K.). Furthermore, E. M. is thankful to the EU Graphene Flagship (CNECT-ICT-604391) and the Swedish Research Council (VR).
\end{acknowledgements}

\appendix
\section*{Appendix}
\subsection{Semi-analytic approach to purely Coulomb-induced CM}

To be able to find optimal excitation conditions and the doping regime for (hot) carrier multiplication, we present a semi-analytical 
approach considering the purely Coulomb-induced carrier dynamics within statistical methods.\cite{winzer12-2}

Here, we focus on the carrier multiplication (CM), but the processes of \textit{hot} carrier multiplication can be treated in an analogous way. Our approach is to evaluate the total carrier density $n$ and the energy density ${\cal E}$ of the electronic system:
\begin{align}
 n = \frac{\sigma_s \sigma_v}{L^2} \sum_{\lambda=e,h;\,{\bf k}} \rho_{\bf k}^{\lambda}
 \ \ \text{and} \ \ {\cal E} = \frac{\sigma_s \sigma_v}{L^2} \sum_{\lambda{\bf k}} \varepsilon_{\bf k}^\lambda \rho_{\bf k}^{\lambda}.
\end{align}
Assuming a Fermi distribution with $\varepsilon_F \neq 0$ for the carrier occupation $\rho_{\bf k}^\lambda$, we obtain the carrier and energy density as a function of the temperature $T$ and the Fermi level $\varepsilon_F$:
\begin{align}
\label{nsupl}n(T, \varepsilon_F)&= \frac{\sigma_s \sigma_v}{2\pi} \sum_{\sigma=\pm} \int_0^\infty \mathrm{d}k \frac{k}{e^{(kv_F+\sigma \varepsilon_F) /(k_B T)} + 1} 
\\\nonumber &= -\frac{2k_B^2}{\pi v_F^2}T^2 \sum_{\sigma=\pm} Li_2(-e^{\sigma \varepsilon_F / T k_B}),
\\[12pt]
\label{Esupl}{\cal E}(T, \varepsilon_F)&=  \frac{\sigma_s \sigma_v v_F}{2\pi} \sum_{\sigma=\pm} \int_0^\infty \mathrm{d}k \frac{k^2}{e^{(kv_F+\sigma \varepsilon_F) /(k_B T)} + 1} 
\\\nonumber&= -\frac{4k_B^3}{\pi v_F^2}T^3 \sum_{\sigma=\pm} Li_3(-e^{\sigma \varepsilon_F / T k_B}),
\end{align}
where $Li_n(.)$ denotes the polylogarithm function.
\begin{figure}[t!]
  \begin{center}
\includegraphics[width=\linewidth]{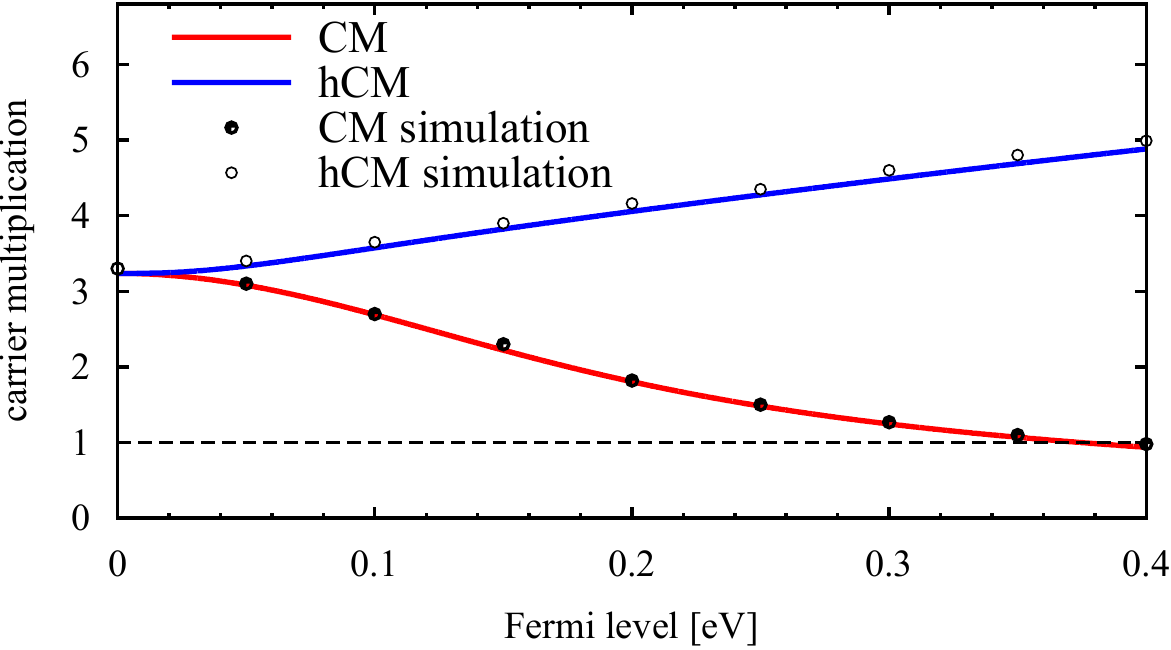}
  \end{center}
  \caption{Analytically obtained CM and hCM as a function of doping at a fixed pump fluence (solid line) of $\varepsilon_{\text{abs}}=\unit[0.1]{\mu J cm^{-2}}$  compared to the numerical 
  solution (dots) obtained by evaluating the graphene Bloch equations. } \label{S1} 
\end{figure}

After an optical excitation, the electronic system is described by the carrier density $n_{exc}=n_0+n_{\text{opt}}$ and the energy density ${\cal E}_{exc}={\cal E}_0+{\cal E}_{\text{abs}}$. Here, ${\cal E}_{\text{abs}}$ denotes the absorbed pump fluence and $n_{opt}=2{\cal E}_{\text{abs}}/\hbar\omega$ the optically excited carrier density with the photon energy $\hbar\omega$. Furthermore, the thermal densities $n_0=n(T_0,\varepsilon_F)$ and ${\cal E}_0={\cal E}(T_0,\varepsilon_F)$ are determined by the initial temperature $T_0$ and Fermi level $\varepsilon_F$. To access the CM factor we need to know the final state of the carrier system after the Coulomb-induced thermalization. We exploit the fact that the Coulomb dynamics conserves the total energy density. Due to Auger processes, the total carrier density is not conserved, but instead the Fermi level. Using Eq. (\ref{Esupl}), we are able extract the final temperature $T_f$ of the hot thermalized system by numerical variation with the condition that $\mathcal{E}(T_f,\varepsilon_F)={\cal E}_{exc}$. Having the final temperature, we can also access the final carrier density $n_f=n(T_f,\varepsilon_F)$ and therewith also the CM factor, cf. Eq. (\ref{eq:cmdef}). For undoped graphene, the purely Coulomb-induced $T_f$, $n_f$, and $CM$ can be obtained analytically. \cite{winzer12-2}   Note however that the approach does not include carrier-phonon scattering, 
which cannot be treated statically. Nevertheless, Fig. \ref{S1} shows an excellent agreement between the analytically [cf. Eqs. (\ref{nsupl})-(\ref{Esupl})] and numerically [cf. Eqs. (\ref{eq:pol})-(\ref{eq:nq})] 
obtained doping-dependent carrier multiplication at a fixed pump fluence of $\varepsilon_{\text{abs}}=\unit[0.1]{\mu J cm^{-2}}$. 
The higher the Fermi level, the smaller is CM, and the larger is hCM. For highly doped graphene samples with $\varepsilon_F>\unit[0.4]{eV}$, CM becomes smaller than $1$. In contrast, hCM reaches values of almost $5$. The same qualitative dependence on doping is found by numerically evaluating the graphene Bloch equations (shown in the main part), which also include the carrier-phonon
coupling. This demonstrates that the presented semi-analytical approach already sufficiently covers the most important aspects of doping-dependent carrier multiplication. For quantitative insights and time-dependent CM, 
the full microscopic approach is required.

\begin{figure}[t!]
  \begin{center}
\includegraphics[width=\linewidth]{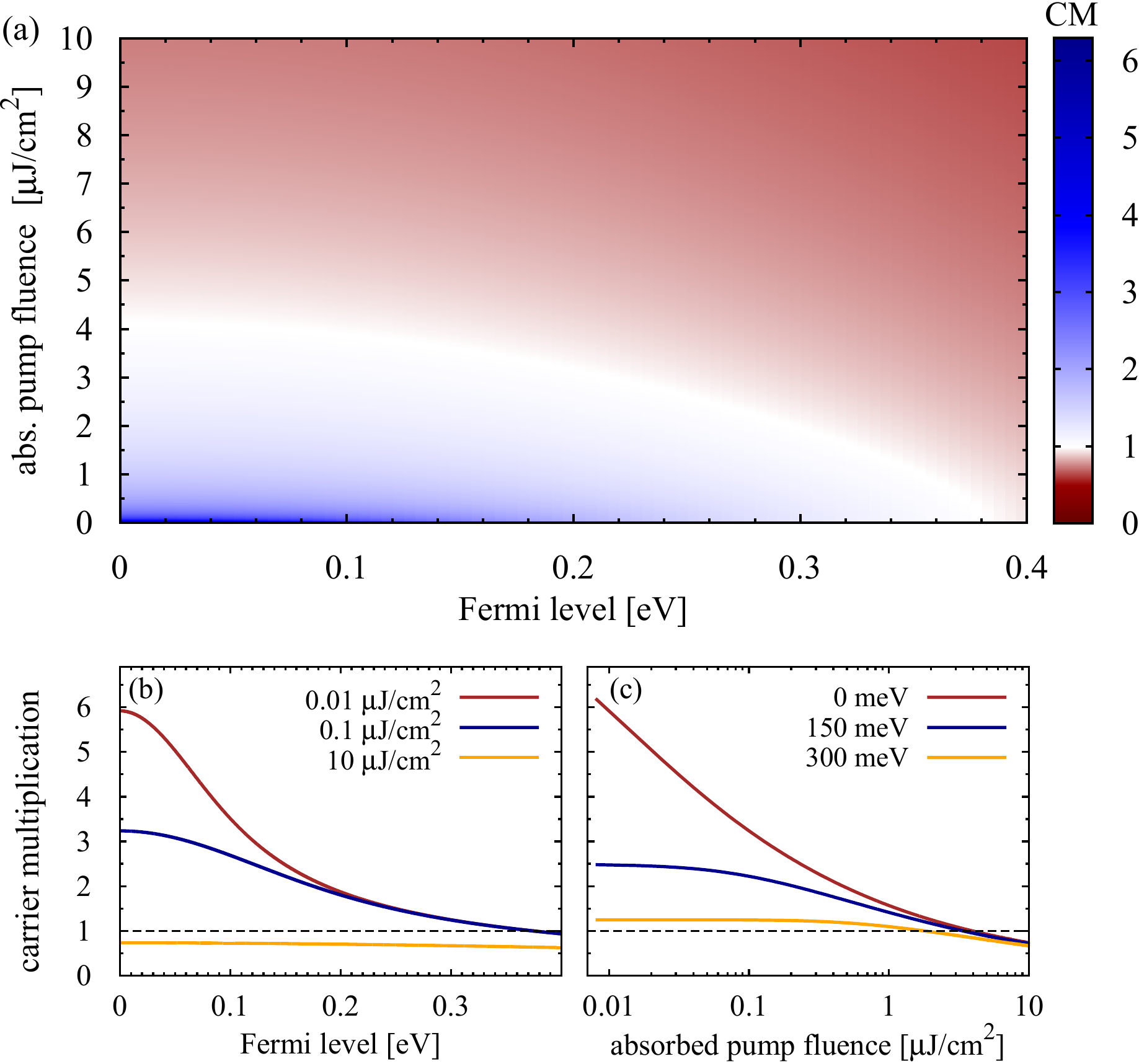}
  \end{center}
  \caption{(a) Carrier multiplication (CM) as a function of Fermi level and pump fluence at room temperature and at an excitation energy of \unit[1.5]{eV}. 
  (b) Doping-dependent CM for three representative pump fluences and (c) fluence-dependent CM for three fixed Fermi levels.} \label{S2} 
\end{figure}

\subsection{Fluence dependence of carrier multiplication}
\begin{figure}[t!]
  \begin{center}
\includegraphics[width=\linewidth]{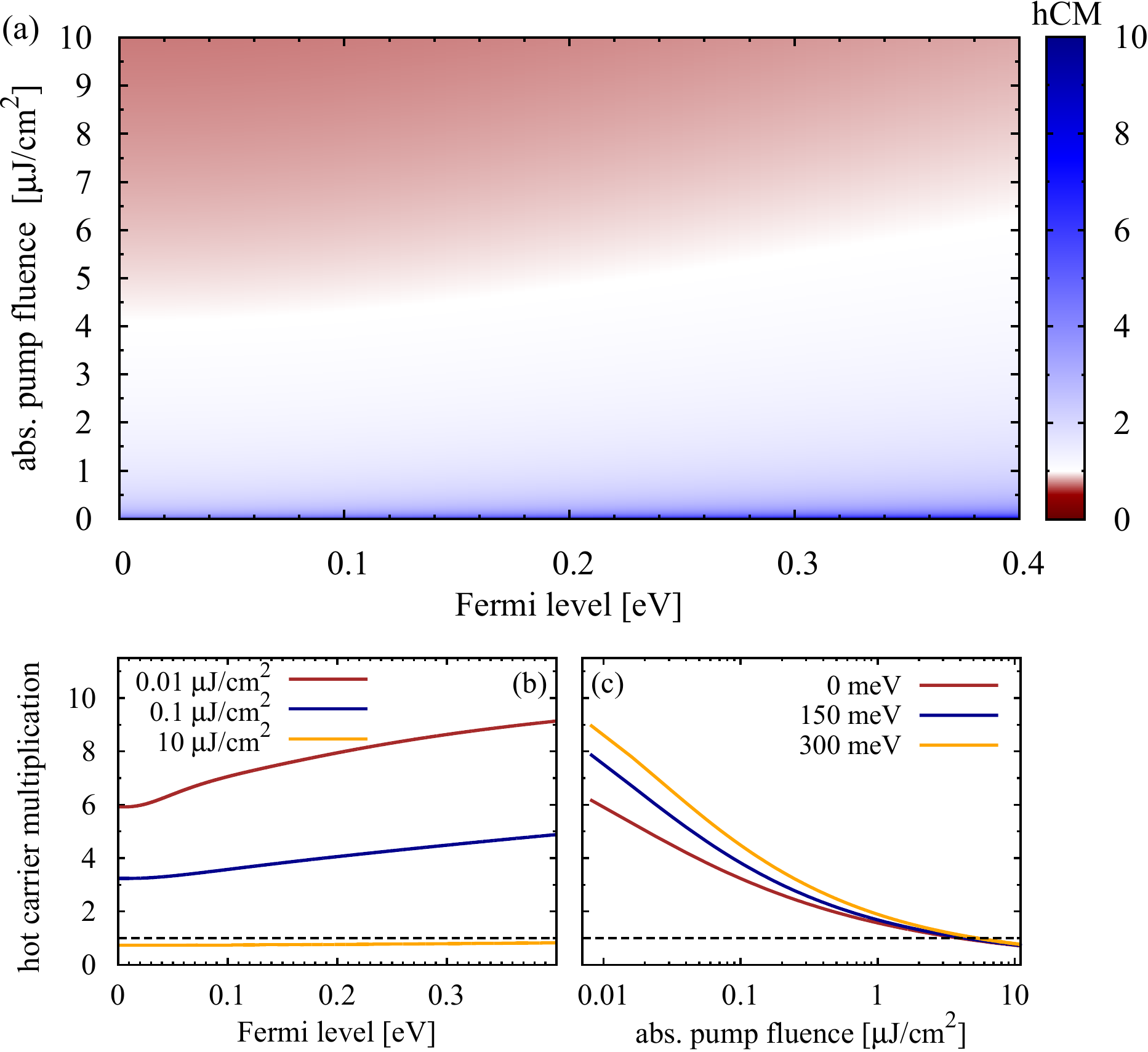}
  \end{center}
  \caption{The same as in Fig. \ref{S2}, however illustrating doping and fluence dependence of the hot carrier multiplication (hCM).} \label{S3} 
\end{figure}

Now, based on the semi-analytical approach presented in the last section,  we discuss the impact of the excitation strength on the appearance of the CM and hCM. 
Figures \ref{S2}(a) and \ref{S3}(a) show surface plots illustrating  doping and  fluence dependence of CM and hCM  at \unit[300]{K} and at an excitation energy of \unit[1.5]{eV}, respectively. 
Our calculations demonstrate that both CM and hCM are strongly suppressed in the strong excitation regime. For pump fluences higher than
approximately $\unit[5]{\mu J cm^{-2}}$ (hot) carrier multiplication does not occur any more. The larger the pump fluence, the more scattering partners are available giving rise to a faster carrier dynamics.\cite{winzer10} 
As a result, the thermalized distribution is reached within the first tens of fs and the asymmetry of impact excitation and Auger recombination rates vanishes much faster. As a result, carrier multiplication clearly decreases at enhanced pump fluences, cf. Figs. \ref{S2}(c) and \ref{S3}(c)
Furthermore, Figs. \ref{S2}(b) and \ref{S3}(b) also illustrate the decrease (increase) of CM (hCM) with the Fermi level, as already discussed in the main part.

\subsection{Dynamical screening}

The efficiency of Auger processes is sensitive to the dynamical screening of the Coulomb interaction that is expressed by the Lindhard equation:\cite{haug04}
\begin{align}
 \label{eq:sc} \varepsilon(\qq,\omega)=1-2V_\qq \sum_{\lambda\lambda'\kk} \frac{\rho_{\kk+\qq}^{\lambda'}-\rho_\kk^\lambda}{\varepsilon_{\kk+\qq}^{\lambda'}-\varepsilon_\kk^\lambda+\hbar\omega+i\delta^\lambda_{\kk,\qq}} \Gamma_{\kk,\qq}^{\lambda\lambda'},
\end{align}
where $\omega$ corresponds to the energy transfer of the scattering processes and the factor 
{$\Gamma_{\kk,\qq}^{\lambda\lambda'}=|\int d^3r \Psi_{\kk}^{\lambda*}({\bf r})e^{i\qq\cdot{\bf r}}\Psi_{\kk-\qq}^{\lambda'}({\bf r})|^2$} includes the 
tight-binding wave functions $\Psi_{\kk}^{\lambda}({\bf r})$.

The appearing momentum-dependent dephasing $\delta^\lambda_{\kk,\qq}$ can be derived from the dynamics of inhomogeneous charge fluctuations
 $
 \braket{\tilde{\rho}_\qq}=-\frac{|e|}{L^3}\sum_{\kk}\braket{a^\dagger_{\kk-\qq}a_\kk}=-\frac{|e|}{L^3}\sum_{\kk}\sigma_{\kk-\qq,\kk},
$. 
The corresponding equation of motion reads:\cite{haug04}
\begin{align}
\nonumber
  \frac{d}{dt} \sigma_{\kk-\qq,\kk} = \frac i \hbar (\varepsilon_{\kk-\qq}-\varepsilon_\kk)\sigma_{\kk-\qq,\kk}+\frac{iV_\qq}{\hbar}(\rho_\kk-\rho_{\kk-\qq})\sum_{\kk'}\sigma_{\kk'-\qq,\kk'}\label{eq:scr}.
\end{align}
Using the ansatz $\sigma_{\kk-\qq,\kk}(t)=e^{-i(\omega+i\delta)t}\sigma_{\kk-\qq,\kk}(0)$, a solution for $\sigma_{\kk-\qq,\kk}$ can be obtained and 
 leads to the dynamical screening expressed by the Lindhard formula with a fluence-dependent dephasing $\delta^\lambda_{\kk,\qq}$.
To obtain the momentum-dependent $\delta^\lambda_{\kk,\qq}$, we write the in-coherent equation of motion for the electron charge density in second-order Born-Markov approximation:\cite{haug04}
\begin{align}
 \frac{d}{dt} \sigma_{\kk-\qq,\kk}|_{nc} &= -\frac \pi \hbar \sum_{ABCD} \tilde{V}^{AB}_{CD} \delta \left(\varepsilon_D+\varepsilon_C-\varepsilon_B-\varepsilon_A\right) \times\\
 &\nonumber\Bigl[\sum_{bcd}V^{\kk b}_{cd}(\sigma^{\phantom{+}}_{\kk-\qq,D}\sigma^{\phantom{+}}_{bC}\sigma_{Ad}^+ \sigma_{Bc}^+ -\sigma^{\phantom{+}}_{Ad}\sigma^{\phantom{+}}_{Bc}\sigma_{\kk-\qq,D}^+ \sigma_{bC}^+) 
\\\nonumber
-&\sum_{abd}V^{\ a \ \ b}_{\kk-\qq d}(\sigma^{\phantom{+}}_{aD}\sigma^{\phantom{+}}_{bC}\sigma_{Ad}^+ \sigma_{B\kk}^+ -\sigma^{\phantom{+}}_{Ad}\sigma^{\phantom{+}}_{B \kk}\sigma_{aD}^+ \sigma_{bC}^+) \Bigr].
\end{align}
This equation contains the momentum-dependent dephasing term, which is given by $\frac{d}{dt} \sigma_{\kk-\qq,\kk}|_{nc} \propto \delta_{\kk,\qq}\sigma_{\kk-\qq,\kk}$.
Neglecting all polarization terms $\sigma_{ab} \rightarrow \rho_{a}$, the dephasing $\delta_{\kk,\qq}$ yields:
\begin{align}
 \label{eq:delta1} \delta^\lambda_{\kk,\qq}= S_{\kk,\qq}^{\text{in},\lambda}+S_{\kk,\qq}^{\text{out},\lambda}
\end{align}
with the in- and out-scattering rates 
\begin{align}
 \nonumber S_{\kk,\qq}^{\text{in},\lambda}=& \frac \pi \hbar \sum_{{\bf a} {\bf b} {\bf c}} \tilde{V}_{\kk, \qq, \bf a, \bf b,\bf c}\, (1-\rho_{\bf a})\rho_{\bf b}\rho_{\bf c} \delta(\varepsilon^\lambda_{\kk}+\varepsilon_{\bf a}+\varepsilon_{\bf b}+\varepsilon_{\bf c}), \\[7pt]
\label{eq:delta2} 
  \nonumber S_{\kk,\qq}^{\text{out},\lambda}=&\frac \pi \hbar \sum_{{\bf a} {\bf b} {\bf c}} \tilde{V}_{\kk, \qq, \bf a, \bf b,\bf c}\,  \rho_{\bf a}(1-\rho_{\bf b})(1-\rho_{\bf c}) \delta(\varepsilon^\lambda_{\kk}+\varepsilon_{\bf a}+\varepsilon_{\bf b}+\varepsilon_{\bf c})
\end{align}
with
$\tilde{V}_{\kk, \qq, \bf a, \bf b,\bf c}=V_{\kk_a \lambda_a \kk_b \lambda_b}^{\kk \ \lambda \ \kk_c \lambda_c}\tilde{V}^{\kk_a \lambda_a \kk_b \lambda_b}_{\kk \ \ \ \lambda \ \kk_c \lambda_c}+V_{\kk_a \lambda_a \kk_b \lambda_b}^{\kk-\qq \lambda \kk_c \lambda_c}\tilde{V}^{\kk_a \lambda_a \kk_b \lambda_b}_{\kk-\qq \lambda \kk_c \lambda_c}$
that also directly appear in the graphene Bloch equations, cf. Eq. (\ref{eq:rhoe}). 
Calculating these scattering rates, we have also a consistent access to the dephasing term influencing the dynamical screening of the Coulomb interaction. 
Note that a vanishing dephasing would suppress the parallel Auger scattering contributions in graphene due to a diverging $\varepsilon(\qq,\omega)$.\cite{brida13}
However, in realistic cases the many-particle processes lead to a significant  dephasing  opening up the Auger channels for the carrier relaxation.\cite{winzer13-1}

\end{document}